# Characterization of highly crystalline lead iodide nanosheets prepared by room-temperature solution processing

Riccardo Frisenda*[1], Joshua O. Island[2,+], Jose L. Lado[3], Emerson Giovanelli[1], Patricia Gant[1], Philipp Nagler[4], Sebastian Bange[4], John M. Lupton[4], Christian Schüller[4], Aday Molina-Mendoza[5,++], Lucia Aballe[6], Michael Foerster[6], Tobias Korn[4], Miguel Angel Niño[1], David Perez de Lara[1], Emilio M. Pérez[1], Joaquín Fernandéz-Rossier[3] and Andres Castellanos-Gomez*[7]

[1] Instituto Madrileño de Estudios Avanzados en Nanociencia (IMDEA-Nanociencia), Campus de Cantoblanco, E-28049 Madrid, Spain.

[2] Kavli Institute of Nanoscience, Delft University of Technology, Lorentzweg 1, 2628 CJ Delft, The Netherlands.

[3] International Iberian Nanotechnology Laboratory (INL), Av. Mestre Jose Veiga, 4715-330, Braga, Portugal.

[4] Department of Physics, University of Regensburg, Universitätsstrasse, Regensburg D-93040, Germany.

[5] Departamento de Física de la Materia Condensada, Universidad Autónoma de Madrid, Campus de Cantoblanco, E-28049, Madrid, Spain.

[6] ALBA Synchrotron Light Facility, Carrer de la Llum 2-26, Cerdanyola del Vallés, Barcelona 08290, Spain.

[7] Instituto de Ciencia de Materiales de Madrid (ICMM-CSIC), Campus de Cantoblanco, E-28049 Madrid, Spain.

[+] Present address: Department of Physics, University of California, Santa Barbara CA 93106 USA.

[++] Present address: Institute of Photonics, Vienna University of Technology, Gusshausstrasse 27-29, A-1040 Vienna, Austria.

*E-mail: riccardo.frisenda@imdea.org; andres.castellanos@csic.es.

**ABSTRACT:** Two-dimensional semiconducting materials are particularly appealing for many applications. Although theory predicts a large number of two-dimensional materials, experimentally only a few of these materials have been identified and characterized comprehensively in the ultrathin limit. Lead iodide, which belongs to the transition metal halides family and has a direct bandgap in the visible spectrum, has been known for a long time and has been well characterized in its bulk form. Nevertheless, studies of this material in the nanometer thickness regime are rather scarce. In this article we demonstrate an easy way to synthesize ultrathin, highly crystalline flakes of $PbI_2$ by precipitation from a solution in water. We thoroughly characterize the produced thin flakes with different techniques ranging from optical and Raman spectros-copy to temperature-dependent photoluminescence and electron microscopy. We compare the results to *ab initio* calculations of the band structure of the material. Finally, we fabricate photodetectors based on $PbI_2$ and study their optoelectronic properties.

**INTRODUCTION**

Shortly after the discovery of graphene by mechanical exfoliation, other two-dimensional (2D) materials were also isolated using the same approach [1-2]. Among the different isolated 2D materials, the family of 2D semiconductors is especially relevant for applications as their intrinsic bandgap makes them suitable for different electronic and optoelectronic devices [3-4]. Although theory predicts a large number of 2D semiconductors that can be produced by mechanical exfoliation (>1000) [5-9], so far the number of compounds that have been experimentally studied is rather limited (<20) [10-12]. The chalcogenides group, and in particular the transition metal dichalcogenides (TMDC) such as molybdenite ($MoS_2$), is probably the most investigated family of 2D materials [13-14]. The transition metal halides (TMH) family, on the other hand, is a scarcely explored family of materials in the few-layer limit, although it contains many examples of layered materials [15-18]. In their bulk form, lead-based TMHs are gaining a great deal of attention as they can be used as a source to synthesize metal halide perovskite solar cells [19-21]. Moreover, ultrathin TMHs present several characteristics that make them an interesting complement to TMDCs, such as their direct bandgap in multilayer samples and the absolute value of the bandgap that spans a range not covered by the TMDCs.

In this study, we use solution-based synthesis of lead iodide ($PbI_2$) followed by mechanical exfoliation to produce thin flakes. We study the isolated layers by micro-reflectance and transmission spectroscopy, atomic force microscopy (AFM), transmission electron microscopy (TEM), Raman spectroscopy, micro X-ray photoemission spectroscopy, temperature-dependent photoluminescence spectroscopy and photocurrent spectroscopy. To complement the material characterization we perform *ab initio* calculations of the thickness dependency of the $PbI_2$ band structure. Even though many methods have been demonstrated to synthesize $PbI_2$ [22], such as the Bridgman method, solution processing and vapor deposition studies of this material are still rather scarce [23]. Our results demonstrate that crystalline large-area ultra-thin lead iodide flakes can be easily synthesized at room temperature from an aqueous solution.

**RESULTS AND DISCUSSIONS**

Lead iodide is a layered material, belonging to the TMHs family, which has been studied in applications such as source for perovskite solar cells [19], photodetectors [24] and nuclear radiation detectors [25]. Its structure, schematically depicted in Figure 1a for a single-layer, is formed by covalently bonded I and Pb atoms arranged in an in-plane hexagonal lattice with out-of-plane van der Waals interactions between the different layers [26]. In order to gain insight into the expected material electronic properties, we computed the band structure based on the crystal structure described above (see computational methods section in the Supplementary Information). Figure 1b shows the band structure calculated for a $PbI_2$ single-layer (left) and for a bulk crystal (right). When going from bulk to a single-layer the nature of the bandgap changes, passing from direct in bulk (with a value of 2.26 eV) to indirect in the monolayer (2.64 eV). The magnitude of the direct gap increases monotonically when thinning down the material from a bulk crystal to a single-layer, as can be seen in Figure 1c and the crossover from a direct to an indirect bandgap happens between one and two layers, in agreement with literature [27].

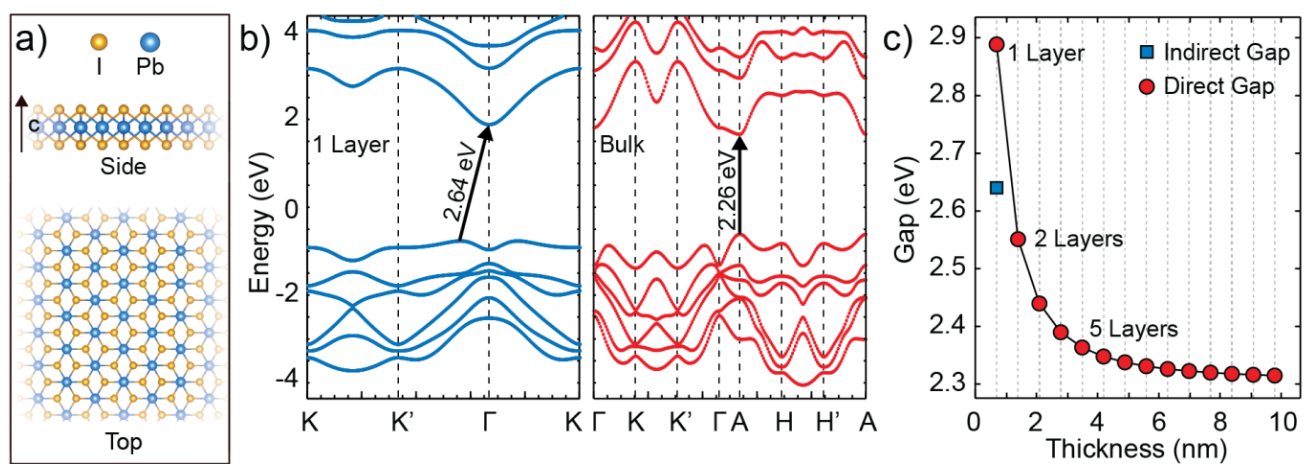

**Figure 1**: a) Crystal structure of a single-layer $PbI_2$. The blue balls are lead (Pb) atoms and the orange balls represent iodine (I) atoms. b) Band structure of a $PbI_2$ single-layer (left) and bulk (right). c) Calculated bandgap energy of $PbI_2$ as a function of the thickness.

Thin $PbI_2$ crystals, with lateral dimensions of up to approximately 100 µm and thicknesses ranging from a few nanometers up to hundreds of nanometers, can be synthesized with a simple precipitation process [23]. We use as a starting material $PbI_2$ in powder (99% purity, Sigma-Aldrich). The solubility of $PbI_2$ in water strongly depends on the temperature in the range between 20 °C and 100 °C, with $PbI_2$ being more soluble in hot water (4.1 g/L at 100 °C) than in cold water (0.8 g/L at 20 °C) [28-29]. Panels 1 and 2 of Figure 2a show a vial containing a supersaturated solution of $PbI_2$ powder (≈0.1 g) in Milli-Q water (20 mL) respectively at 25 °C and at 100 °C. The change in solubility of $PbI_2$ is reflected by the change in the color of the liquid. During the cooling down of the solution, dissolved Pb and I species start to aggregate and form crystals of $PbI_2$, which appear as a fine dust in panels 3-5 of Figure 2a (see the comparison between the top and bottom rightmost panels in Fig. 2a). To isolate the $PbI_2$ nanosheets, a 1 mL droplet of the saturated $PbI_2$ solution at 100 °C is drop-cast onto the surface of a polydimethylsiloxane (PDMS) stamp. When the drop cools down, $PbI_2$ crystals of various sizes and thicknesses grow, as shown in the optical micrographs recorded at different times shown in Figure 2b. The crystallization process can last from a few seconds to several minutes and by monitoring the process with a microscope (20x magnification, working distance of 20.5 mm) we can draw up the liquid at the desired time from the PDMS surface with a pipette in order to reach the desired density of crystals on the PDMS. The advantage of this method is that the as-produced crystals can be readily transferred to another substrate or onto pre-patterned nanoscale structures using the deterministic transfer method [30]. Moreover, thinner flakes can be achieved by mechanical exfoliation of the as-crystallized flakes with another PDMS stamp. The crystallization process is reproducible and gives rise to $PbI_2$ crystallites with hexagonal shape (see Supplementary Information), consistent with previous studies using different $PbI_2$ syntheses [27, 31-32].

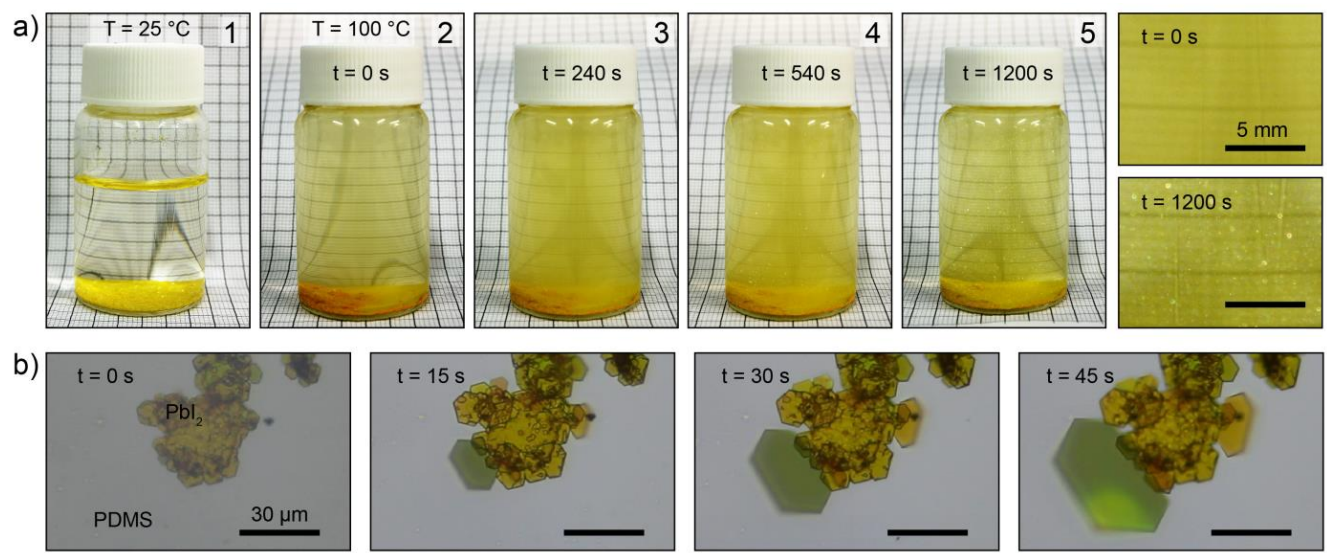

**Figure 2**: a) Time sequence of the steps of the synthesis of PbI$_2$ crystallites from solution. First, PbI$_2$ in powder is added to Milli-Q water at 25 °C (1) and heated to 100 °C (2). When the hot solution cools down crystallites of PbI$_2$ form and appear as a fine dust (3-5). The last two panels on the right are zoomed views of the solution from steps 2 and 5. b) Microscope pictures taken at different times of the formation of a PbI$_2$ thin crystal from a drop deposited on a PDMS substrate.

To relate the thickness of the PbI$_2$ crystals to the optical properties of the material we first characterized the PbI$_2$ flakes by a combination of optical and atomic force microscopy (AFM). Figure 3a-e shows optical micrographs of different flakes that have been transferred onto a SiO$_2$/Si substrate with 285 nm oxide. Similar to the case of graphene and other 2D materials, different flakes deposited on SiO$_2$/Si present different colors according to their thickness because of thin-film interference effects [33]. We have used AFM measurements to accurately determine the thickness of the flakes in Figure 3. From comparison of the AFM data to the microscope pictures we extract a calibration chart, shown in Figure 3f, which can be used to estimate the thickness of PbI$_2$ flakes from their color at first glance. More quantitative information can be obtained from the analysis of the red/green/blue channels of the optical images. Figure 3g shows the optical contrast of the PbI$_2$ flakes on SiO$_2$ calculated for different flakes from the intensity of the red/green/blue channels according to $C=(I_{R,flake} - I_{R,sub})/(I_{R,flake} + I_{R,sub})$, where $C$ is the optical contrast and $I_{R,flake}$ ($I_{R,sub}$) is the intensity of the light reflected by the PbI$_2$ flake (substrate). When the contrast is zero the flake is not detectable, while when the contrast is positive (negative) the flake appears brighter (darker) than the substrate. In all the three channels the optical contrast of PbI$_2$ flakes does not present a monotonic dependence on the thickness, but instead it shows oscillations and it explores both positive and negative values because of thin film interference effects. Notice that the PbI$_2$ flakes in panels d, e and g of Figure 3 have been mechanically exfoliated from thicker flakes.

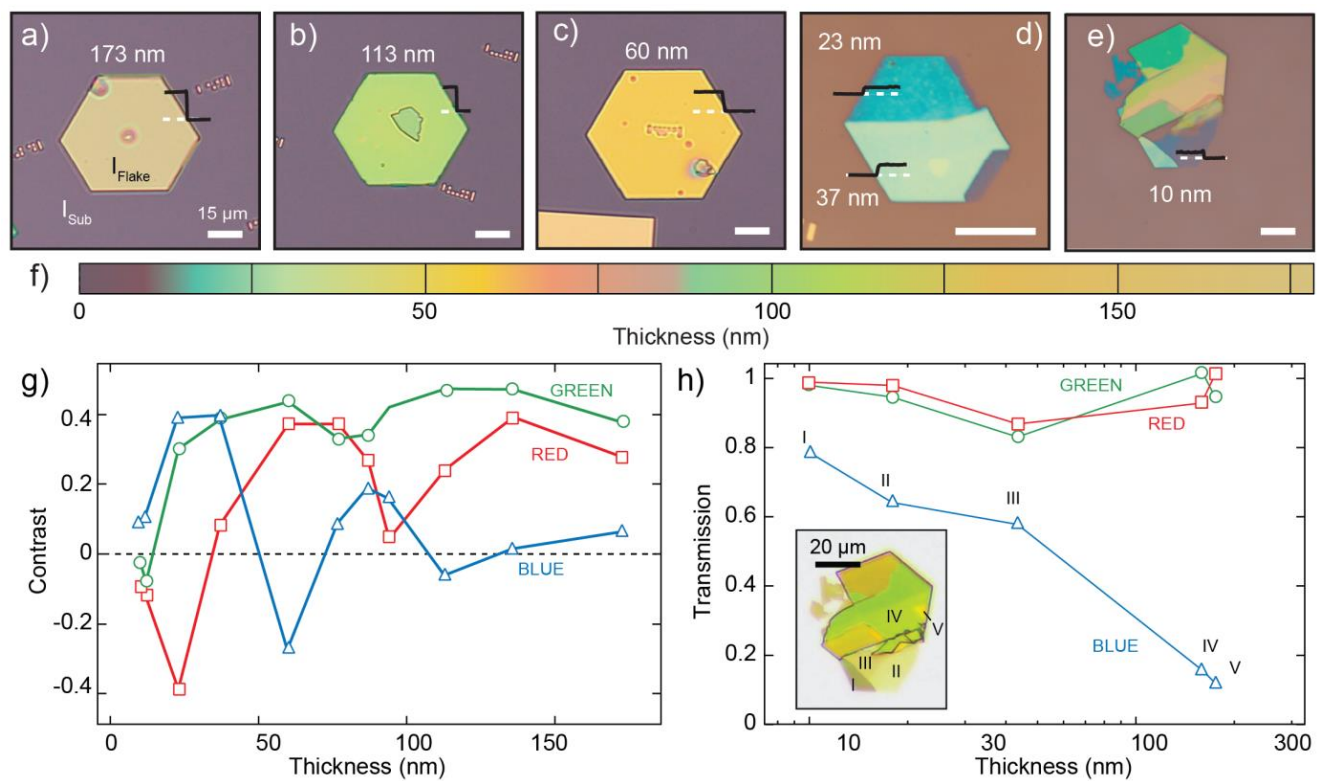

**Figure 3**: a-e) Optical image of a PbI$_2$ flake transferred onto a 285nm SiO$_2$/Si substrate. The black line is the topographic profile measured with an AFM to extract the flake thickness. f) Color chart for thickness calibration of PbI$_2$ flakes on SiO$_2$/Si. l) Optical contrast of the flakes as a function of the thickness calculated from the red, green and blue channels of panels (a)-(e). h) Transmission of PbI$_2$ on PDMS as a function of the flake thickness. Inset: optical micrograph of a PbI$_2$ flake showing various regions of different thickness recorded in transmission illumination mode (white light source). The roman numerals indicate the regions of the flake used to extract the transmission and the thickness data.

Another simple method that is used to estimate the thickness of ultra-thin 2D materials is the analysis of a photograph of the flake under study deposited on a transparent substrate and recorded in transmission illumination mode with a white light source. The inset of Figure 3h shows a microscope picture of a PbI$_2$ flake on PDMS illuminated in transmission mode. Different colors and shades correspond to different values of thickness. We extract the transmission, T = $I_{T,flake}/I_{T,sub}$, for red/green/blue channels at five different points on the flake. After recording the transmission illumination mode microscope optical photograph we transfer the flake onto a SiO$_2$/Si substrate to perform AFM measurements and extract the thickness of the flake. Figure 3h shows the local transmission of the flake as a function of the thickness for red/green/blue channels. Due to the large band gap of the PbI$_2$, the transmission of the flakes remains close to unity in the red and green channels even for thicker flakes. By contrast, the transmission extracted from the blue channel shows a monotonic dependence on the thickness and it can be used to have a rough and quick estimation of the thickness of ultra-thin PbI$_2$ crystals.

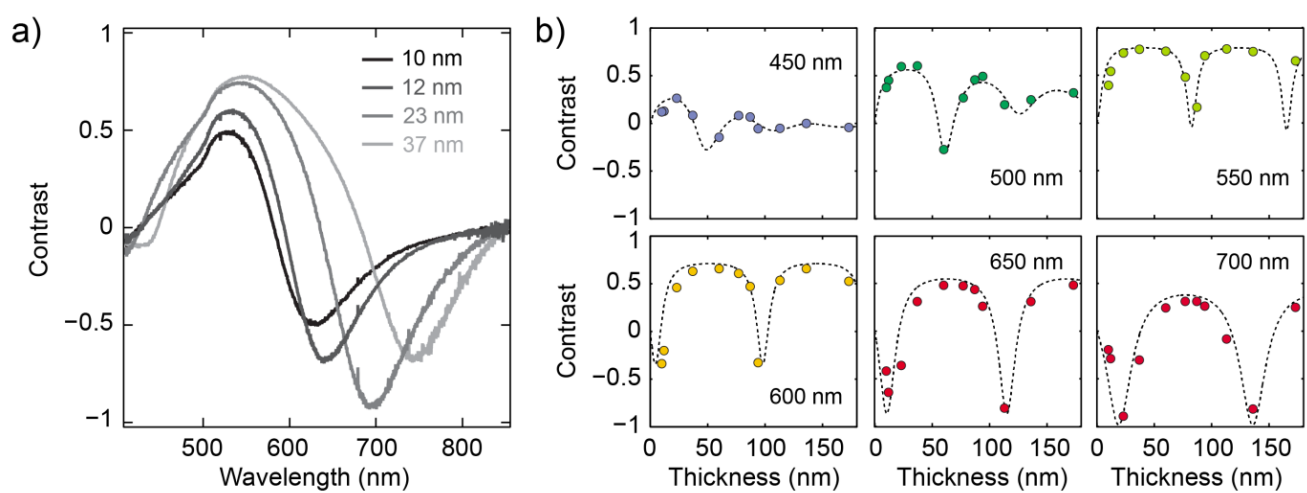

**Figure 4**: a) Wavelength-resolved optical contrast of PbI$_2$ flakes of different thickness. b) Optical contrast as a function of thickness extracted at different wavelengths. The dashed lines are fits to the Fresnel equation.

To extract more quantitative information about the optical properties of PbI$_2$ we study the thickness dependence of the optical contrast of PbI$_2$ on SiO$_2$/Si resolved in energy. To this end, we employ a micro-reflectance spectroscopy set-up that was previously described elsewhere [34-35]. Briefly, the flakes are illuminated in normal incidence with the white light produced by a tungsten halogen lamp, and the reflected light is collected with an optical fiber (core diameter of 105 μm), which acts as a confocal pinhole, placed at the image plane of the trinocular of an optical microscope. The other end of the fiber optic cable is connected to a compact CCD spectrometer (Thorlabs), allowing to measure the spectrum of the reflected light from 400 nm to 900 nm with ≈1 nm resolution. By measuring the reflectance spectrum of the bare substrate (SiO$_2$/Si) and that of the PbI$_2$ flake, we can determine the optical contrast $C$ in a broad range of the visible and near-infrared part of the electromagnetic spectrum. Figure 4a shows the optical contrast spectra rec-orded for four flakes with different thickness. A comparison of the different spectra shows that the largest positive contrast can be achieved at a wavelength of ≈550 nm, independent of the $PbI_2$ thickness. At longer wavelengths the optical contrast spectra reverse their sign, going from positive to negative, and present a dip with the largest negative contrast at a wavelength in the range between 600 nm and 800 nm depending on the thickness of $PbI_2$. Figure 4b shows the optical contrast as a function of the $PbI_2$ thickness, extracted at different wavelengths between 450 nm and 700 nm. The contrast data have an oscillatory behavior with a larger period for longer wavelengths, which can be well fit to the Fresnel equation represented by dashed lines in Figure 4b [33]. From each spectrum taken at a certain wavelength $\lambda$, we extract the complex refractive index at that wavelength $\underline{n}(\lambda)$.

From the Fresnel equation fit of Figure 4b, by fitting hundreds of curves with different wavelengths between 400 nm and 800 nm, we calculate the wavelength-resolved $PbI_2$ complex refractive index $\underline{n} = n - i\kappa$. Figure 5a shows the refractive index $n$ and the extinction coefficient $\kappa$ of $PbI_2$ as a function of the wavelength. The refractive index is a slowly varying function of the wavelength while the extinction coefficient, which is related to the absorption, shows sharper features. For wavelengths longer than 510 nm, $\kappa$ takes values close to zero, indicating that the material does not absorb at these wavelengths. On the other hand, at wavelengths shorter than 510 nm the extinction coefficient steadily increases, which is due to direct band-to-band transitions. Finally, we use the estimated complex refractive index to calculate the optical properties of a thin film of $PbI_2$ under normal incidence using the Fresnel equation. Figure 5b displays a colormap that represents the calculated optical contrast $C$ of a monolayer $PbI_2$ as a function of the illumination wavelength and the thickness of the $SiO_2$ dielectric layer. For a given excitation wavelength, the contrast oscillates around zero with a period that increases with the wavelength of the incident light. The maximum contrast is achieved for $SiO_2$ thicknesses of approximately 70 nm, 220 nm and 350 nm. Notice that the abrupt change in the colormap around 510 nm is due to a change in the magnitude of the optical contrast caused by the strong absorption of $PbI_2$ at the bandgap (see the graph of $\kappa$ vs wavelength).

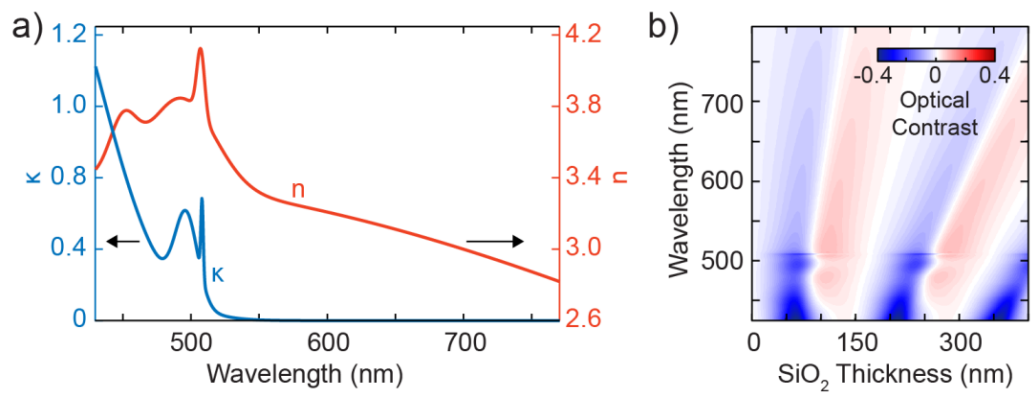

**Figure 5**: a) Real ($n$) and imaginary ($\kappa$) components of the refractive index ($\underline{n} = n - i\kappa$) calculated from the fits of Figure 4b. b) Calculated optical contrast as a function of the illumination wavelength and SiO$_2$ thickness for PbI$_2$ on a SiO$_2$/Si substrate.

Raman spectroscopy, which has been demonstrated to be a very powerful tool to characterize the quality and the thickness of 2D materials [36-37], was employed to characterize the flakes previously studied in Figures 3 and 4. Figure 6a shows the Raman spectra collected for flakes 23 nm to 180 nm thick whose intensities are all normalized by the intensity of the peak at 520 cm$^{-1}$, arising from transverse optical modes of the Si substrate. The excitation wavelength is 532 nm (2.33 eV), which is close to resonance with the fundamental absorption band (≈2.3 eV). The spectra show an intense and broad peak centered at approximately 214 cm$^{-1}$ that is characteristic of the $A_1^2$ vibrational mode of hexagonal PbI$_2$ and is indicative of the crystallinity of the sample [38-40]. The intensity of this peak increases with the PbI$_2$ thickness and can be used to estimate the thickness of the material without the need of topographic measurements. Figure 6b shows the ratio between the $A_1^2$ peak and the Si 520 cm$^{-1}$ peak as a function of the PbI$_2$ thickness. Due to the strong background in the Raman spectra we cannot access the energy range below 150 cm$^{-1}$ which contains additional Raman peaks of PbI$_2$ that are sensitive to the crystalline phase of the material [41-42].

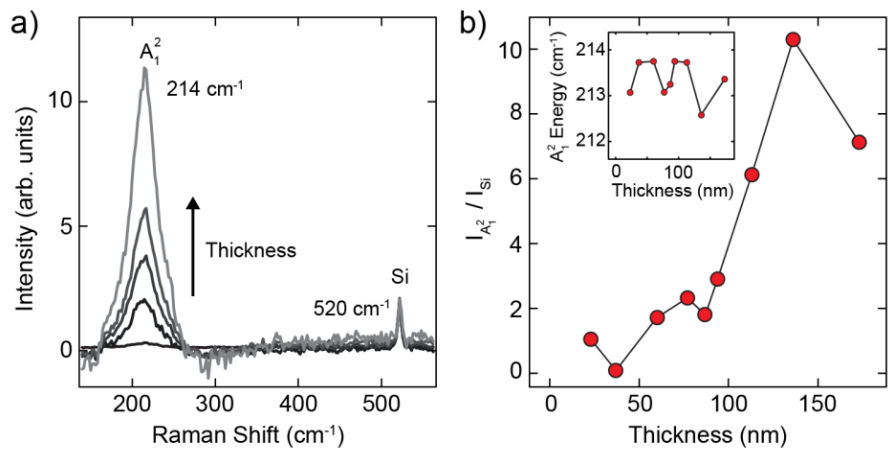

**Figure 6**: a) Raman spectra of $PbI_2$ flakes with different thickness recorded with an excitation wavelength of 532 nm. All the spectra have been normalized by the intensity of the Si peak at 520 cm$^{-1}$. b) Ratio between the intensities of the $A^1_2$ $PbI_2$ peak (214 cm$^{-1}$) and the Si peak (520 cm$^{-1}$) as a function of the $PbI_2$ thickness. Inset: Center of the $A^1_2$ peak as a function of $PbI_2$ thickness.

To characterize the crystal structure of the fabricated $PbI_2$ flakes we study them with high-resolution transmission electron microscopy (TEM). The samples are fabricated by simply drop casting the saturated hot solution of $PbI_2$ and water onto a nickel TEM grid with 6 μm holes, which leads to crystallization of thin hexagonal flakes covering a few holes. Figure 7a shows a high-resolution TEM image (20 nm x 20 nm) of a $PbI_2$ flake where individual Pb atoms are visible and in the inset a low-magnification image (3 μm x 3 μm) of the same flake is shown. To better analyze the atomic arrangement we show a zoom of the TEM image in Figure 7b where we superimposed a drawing of the $PbI_2$ lattice in the 2H polytype. We find a perfect match between the atomic arrangements in the TEM image and the $PbI_2$ 2H polytype generated lattice, which is in agreement with the minimum energy structure found in the *ab initio* calculations and with previous studies with similar synthesis.[26, 43] Notice that because of differences in the notation, the 2H polytype in $PbI_2$ is identical in stacking to the 1T polytype in transition metal dichalcogenides such as $MoS_2$ or $WSe_2$.[44]

Additionally, using a Matlab algorithm we extract the distance between all the pairs of prime- and second-nearest neighbor atoms in Figure 7a (see the Supplementary Information). We find an average distance of (2.65 ± 0.03) Å for the nearest neighbors, which is consistent with the lead-lead distance and corresponds to the lattice spacing of the (100) planes. The second-nearest neighbor pairs of atoms have an average distance of (4.56 ± 0.02) Å which matches perfectly the lattice constant of $PbI_2$ [15]. The two-dimensional Fourier transform of the TEM image, shown in Figure 7c, reveals the long-range order of the imaged lattice. Fi-nally, the clear diffraction spots of the electron diffraction image of Figure 7d are consistent with the single-crystalline nature and long-range order of the fabricated PbI$_2$ flakes.

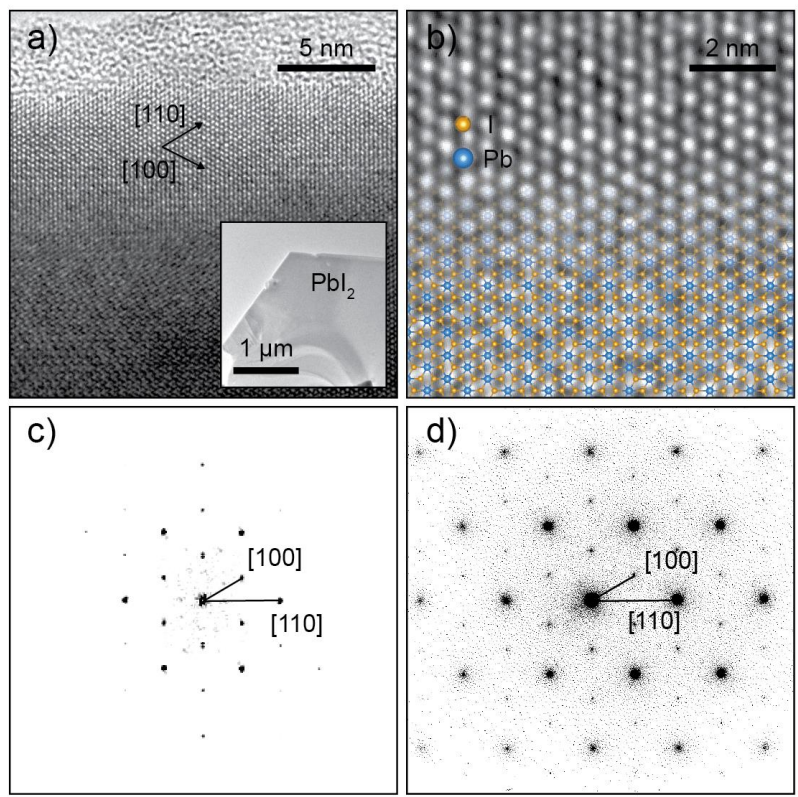

**Figure 7**: a) High-resolution transmission electron microscopy (TEM) image of a PbI$_2$ thin flake. Inset: low magnification TEM image of the flake. b) Zoom in the TEM image of panel (a). Superimposed to the TEM image there is a sketch of 2H-PbI$_2$ crystal lattice. Lead atoms are represented in blue and iodine in orange. c) Two-dimensional Fourier transform of the TEM image of panel (b). d) Selected area electron diffraction of the PbI$_2$ flake.

We also perform micro-XPS and low-energy electron microscopy measurements of PbI$_2$ crystallites transferred onto a platinum substrate. The results, discussed in the Supplementary Information, indicate a steady desorption of iodine with time when the sample is placed in ultrahigh vacuum ($10^{-10}$ mbar) and under X-ray irradiation. We observe a full desorption of the iodine atoms when the samples were annealed at temperatures larger than 150 °C in ultrahigh vacuum conditions.

After establishing the crystallinity of the PbI$_2$ ultra-thin crystals, we investigate their optical properties with temperature-dependent photoluminescence (PL) measurements. The PbI$_2$ flakes grown from solution on PDMS are deterministically transferred onto a SiO$_2$/Si substrate and subsequently studied in a self-built micro-photoluminescence system, details are published elsewhere [45]. Briefly, a continuous-wave laser (central wavelength 375 nm) is coupled into a 100x microscope objective and focused to a sub-micron spot on the sample, which is mounted in a He-flow cryostat. The PL emission from the sample is collected by the same microscope objective and coupled into a spectrometer with a Peltier-cooled charge-coupled device (CCD) sensor.

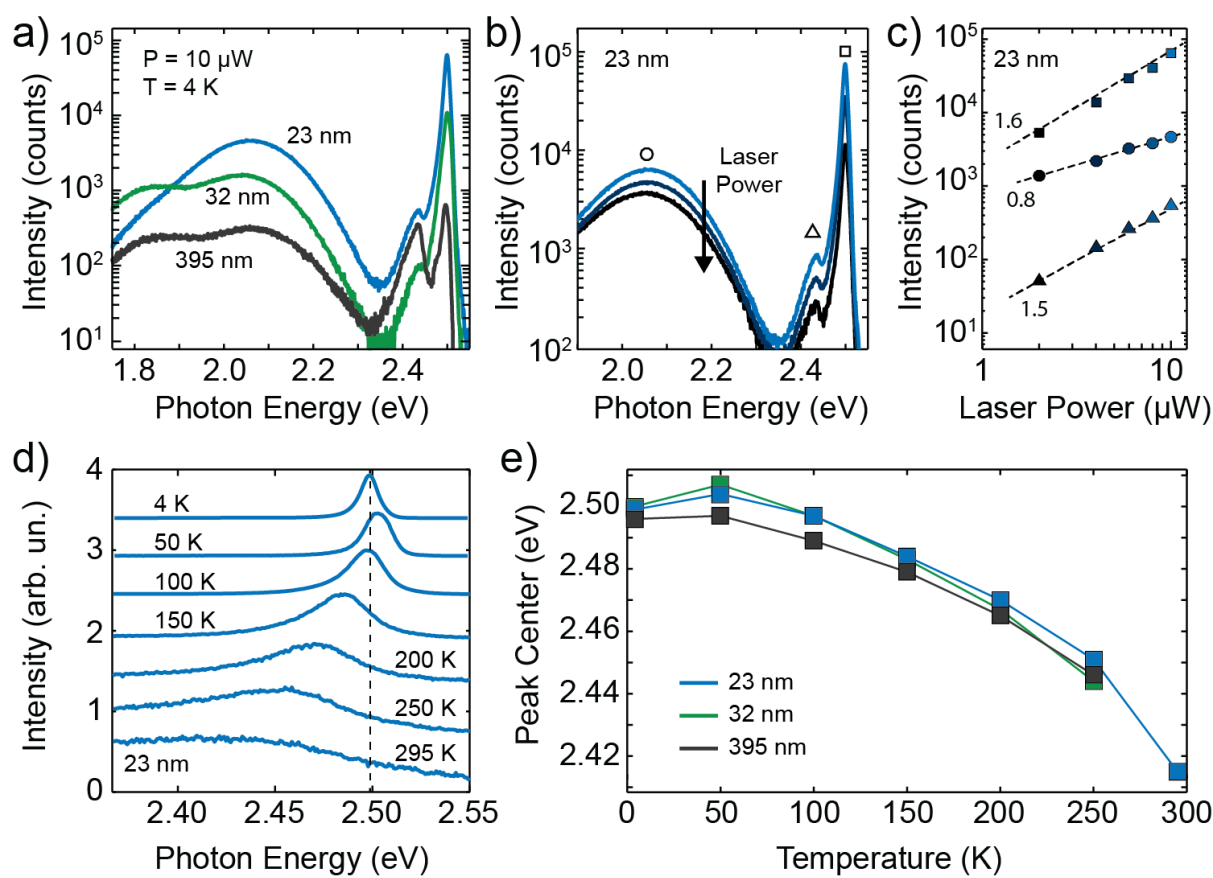

**Figure 8**: a) Photoluminescence spectra of three different PbI$_2$ flakes on SiO$_2$/Si with an excitation wavelength of 375 nm and power of 10 µW. c) Log-log plot of the intensity of the photoluminescence peaks at 2.49 eV (squares), 2.42 eV (triangles) and 2.05 eV (circles) as a function of laser excitation power for the 23 nm thick flake. The numbers indicate the exponent of a power law fit performed on each curve (a value of 1 indicates a linear dependence). d) Photoluminescence spectra of a 23 nm thick PbI$_2$ flake recorded at dif-

ferent temperatures from 4 K to 295 K. e) Energy of the highest-energy photoluminescence peak as a function of temperature.

In Figure 8a we plot the PL spectra of three different flakes with thickness from 23 nm to 395 nm in a semi-logarithmic representation recorded at 4 K and with the same excitation power of 10 µW (see the Supplementary Information for optical pictures of the flakes). The spectra present common features, starting from high energy we find a peak centered at 2.49 eV. The 395 nm thick flake shows a shoulder on the low-energy side of the first peak. Additionally, in all three spectra a weaker peak is present at 2.42 eV and a broad peak or double peak is centered at lower energies between 2.1 eV and 1.8 eV. The peak at 2.49 eV is the fundamental excitonic peak due to direct band-to-band transition, and its energy is in good agreement with the onset of absorption observed in the refractive index data discussed above [46-47]. We note that all flakes investigated in the PL measurements are so thick that neither changes of the band structure nor exciton confinement effects which would shift the transition energies are to be expected. Remarkably, we observe that the thinnest flake (23 nm) shows the highest PL intensity. The peak at 2.42 eV can be interpreted as due to bound exciton by a comparison with literature [40]. On the other hand, the lowest peak at 2.05 eV, which is called G band, does not have an excitonic origin, but is due to defects in the surface of the $PbI_2$ samples and especially to the presence of $Pb^+$ ions [40, 42]. Figure 8b shows the photoluminescence spectra of the 23 nm thick flake recorded at different excitation powers. We extract the intensity of the free and bound excitonic peaks and of the G band for different laser powers and Figure 8c shows the results for the 23 nm thick flake. From the log-log plot one can see that for all the peaks the emission intensity $I$ displays a power-law dependence on the excitation power, which can be described by the formula $I \approx P^k$. The dependency of the PL intensity on the laser power is an indicator of the nature of the different recombination processes. The intensity of the peaks at 2.49 eV and 2.42 eV can be fit with exponents $k$ = 1.6 and 1.5 respectively, indicating a super-linear scaling of I versus P. By contrast, the G band intensity scales sub-linearly with the excitation power with exponent $k$ = 0.8. According to literature, a value of $k$ in the range between 1 and 2 is related to exciton-like transitions while a value of $k$ smaller than 1 is related to recombination of trapped with free carriers [48-49], confirming that the G band is due to defects in the crystal structure and that the higher energy peaks are due to excitonic transitions.

We also investigate the temperature dependence of the PL emission. Figure 8d shows the PL spectra of a 23 nm thick $PbI_2$ flake recorded between 4 K and 295 K. Between 4 K and 50 K, we observe a slight blueshift for the 23 and 32 nm thick flakes (see Fig. 8e), indicative of a de-trapping of excitons bound to shallow de-fects such as charged surface adsorbates. As the temperature is increased further, the main excitonic peak shows an energetic redshift and its intensity decreases. This temperature dependence of the excitonic peak reflects the thermally induced change in the bandgap of PbI$_2$ and the graph of the excitonic peak center versus temperature of Figure 8e can be fit to the Bose-Einstein model, see Supplementary Information.

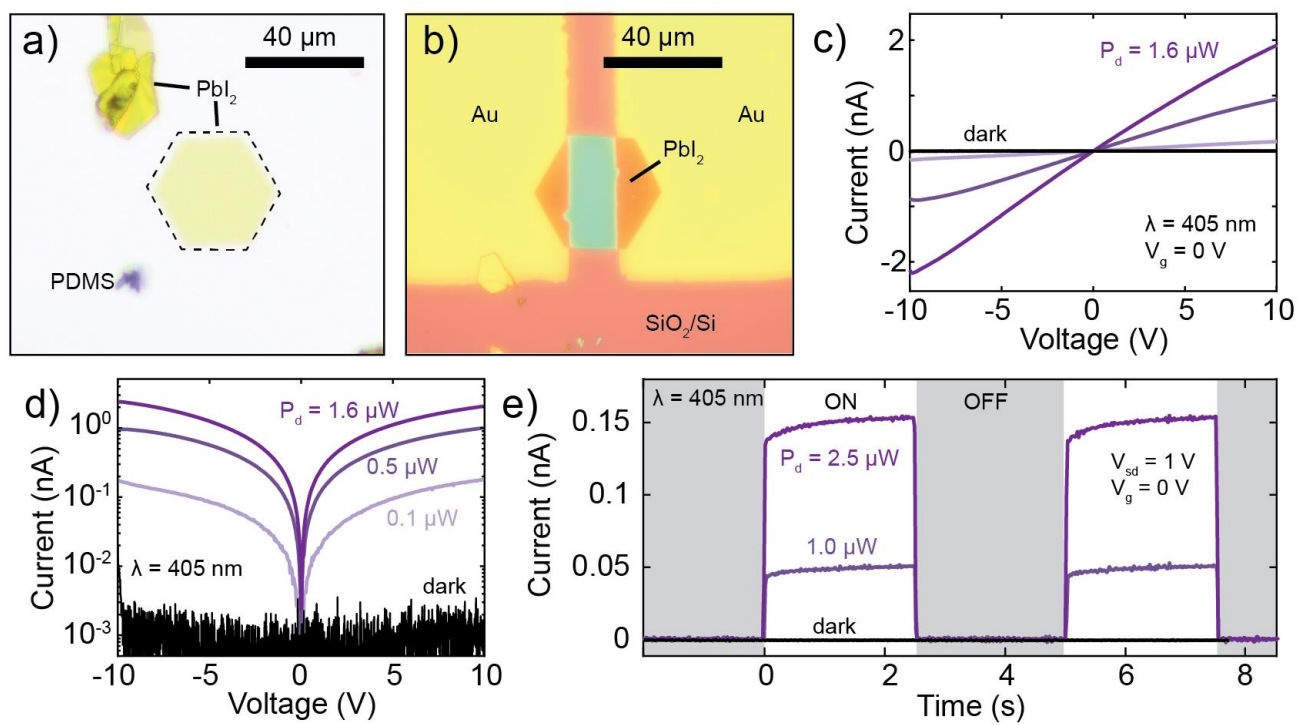

**Figure 9**: a) Microscope picture of a PbI$_2$ flake on PDMS in transmission illumination mode. The dashed line highlights the PbI$_2$ flake used to fabricate the field effect transistor. b) Optical picture of the same flake shown in panel (a) after being transferred between two Au electrodes. c) Current versus source-drain voltage characteristics of the device in dark and under illumination with a 405 nm LED and different powers. d) Semi-logarithmic representations of the $IV_{SD}$ curves of panel (c). Notice that the sign of the current at negative bias has been reversed. e) Current versus time at fixed bias voltage in dark (black curve) and under modulated illumination (violet curves).

Finally, we study the optoelectronic properties of thin PbI$_2$ flakes by fabricating photodetector devices. In total we fabricated five PbI$_2$-based photodetectors, four of which were fabricated on a SiO$_2$/Si substrate while one on a transparent and flexible polycarbonate substrate (see section 6 of the Supplementary Information). We transfer a 15 nm thick PbI$_2$ flake from a PDMS substrate, see Figure 9A, bridging drain and source Au electrodes pre-patterned on a SiO$_2$/Si substrate (see Supplementary Information for the AFM profile of the flake). The Si is heavily doped in order to use it as a back gate electrode and in the measurements discussed below it is kept at zero voltage. An optical picture of the final device is shown in Figure 9b. We record current versus drain-source voltage ($IV_{SD}$) characteristics of the device in dark and under illumination. We use various high-power LEDs, with central wavelengths ranging from 375 nm to 1050 nm, that are focused on the surface of PbI$_2$ on spot diameter of 200 µm providing a homogenous illumination over the entire device area. Figure 9c shows $IV_{SD}$ characteristics acquired in dark and under illumination using a 405 nm LED with increasing optical power. The optical power indicated in the plots, $P_D$, is the power of the radiation falling on the device, calculated by dividing the total beam power by the spot area and multiplying the result for the area of the PbI$_2$ transistor channel (684 µm$^2$) estimated from the microscope picture. The photocurrent (difference between the current upon illumination and in dark) generated in the device displays a sub-linear dependence on the incident power (exponent equal to 0.7, see Supporting Information) which suggests the influence of defects in the dynamics of the charge carriers [50-52]. The semi-logarithmic representation of the $IV_{SD}$ of Figure 9d reveals an on-off ratio larger than 2000 for the largest optical power.

We also study the time response of the device by recording the current at a fixed source-drain voltage as a function of time ($It$) while modulating the light intensity with a ≈0.5 Hz square wave. Figure 9e report the current at $V_{SD}$ = 1 V in dark and under modulated illumination. We cannot directly resolve the switching times of the device when the illumination is turned on or off with our experimental time resolution. Instead, we can define an upper boundary of 20 ms for the device switching time. By recording $It$ traces with different excitation wavelength all set at the same optical power we can study the energy dependence of the PbI$_2$ responsivity. Figure 10a shows three $It$ traces recorded at wavelengths $\lambda$ = 405, 455 and 505 nm. The device displays the same fast response to all the wavelengths and has the largest response for the shortest wavelength. We extract the photocurrent generated at each wavelength $I_{Ph}(\lambda)$ in the on state and we calculate the responsivity according to $R(\lambda) = I_{Ph}(\lambda) / P_D$. In the top panel of Figure 10b we plot the responsivity extracted from the $It$ traces as a function of excitation wavelength. The responsivity is zero for wavelengths larger than 505 nm and it steadily increases at shorter wavelengths with the largest value recorded equal to 1.3 · 10$^{-3}$ A/W, which is a low value compared to other multilayer 2D materials [52]. The behavior of the responsivity is consistent with the bandgap of PbI$_2$ of 2.38 eV, corresponding to a wavelength of 521 nm, extracted from the Tauc plot of the absorbance (with Tauc exponent for a direct gap) shown in the bottom panel of Figure 10b [32]. The different photodetectors built from PbI$_2$ and shown in the supplementary information have comparable responsivities to the main text device.

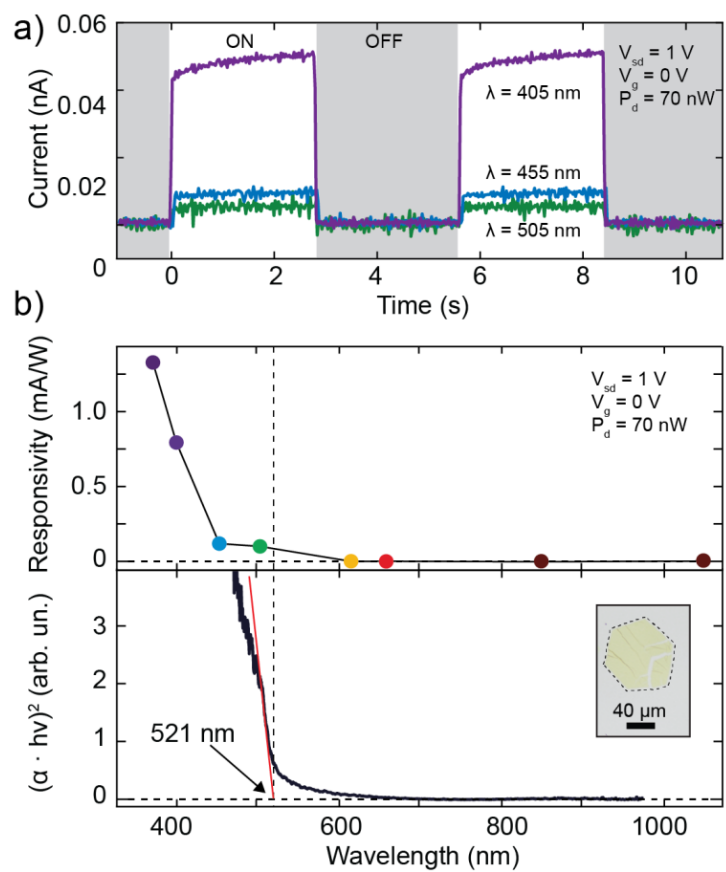

**Figure 10**: a) Current versus time at fixed bias voltage under modulated illumination with different wavelengths and same optical power. b) Top: Responsivity of the PbI$_2$ device for different wavelengths. Bottom: Tauc plot with exponent 2 (direct gap) of the absorbance of the PbI$_2$ shown in the inset, extracted from transmittance measurements. The red line is a linear fit to the data and the energy at which it intersects the x-axis is the estimated bandgap.

**CONCLUSIONS**

In conclusion, we have discussed an easy route to synthesize highly crystalline nanosheets of PbI$_2$ from an aqueous solution and we presented a thorough characterization of the resulting material. We studied the reflectance of PbI$_2$ as a function of the crystal thickness to extract the refractive index components $n$ and $\kappa$ of the material and found that the onset of absorption spectrally matches the main emission peak we observed in photoluminescence. Additionally, temperature- and power-dependent photoluminescence measurements revealed defect-related emission and a redshift of the main emission peak with increasing temperature due to a decreasing band gap. Raman spectroscopy, TEM and, electron diffraction and micro-XPS measurements demonstrated the crystallinity of the as-grown and exfoliated PbI$_2$ samples. Finally we fabricated photodetectors based on PbI$_2$ thin flakes and studied their optoelectronic properties. The experiments have been compared to *ab initio* calculations of the electronic band structure of the material.

## ACKNOWLEDGEMENTS

We acknowledge financial support from the European Commission under the Graphene Flagship (CNECTICT-604391), and European Research Council (ERC-StG-MINT 307609), the MINECO, the Comunidad de Madrid, the Netherlands Organisation for Scientific Research (NWO), and the German Science Foundation (DFG). We are also grateful to B. Galler and R. Zeisel (OSRAM OS, Regensburg) for technical assistance in the photoluminescence experiments. J.L.L. and J.F.R. acknowledge financial support by Marie-Curie-ITN 607904-SPINOGRAPH. J.F.R. acknowledges financial support from MEC-Spain (MAT2016-78625-C2).

## COMPETING INTERESTS

The authors declare no competing financial interests.

## FUNDING

A.C.G. European Commission under the Graphene Flagship: contract CNECTICT-604391

A.C.G. MINECO: Ramón y Cajal 2014 program RYC-2014- 01406

A.C.G. MINECO: program MAT2014-58399-JIN

A.C.G. Comunidad de Madrid: MAD2D-CM program (S2013/MIT-3007)

R.F. Netherlands Organisation for Scientific Research (NWO): Rubicon 680-50-1515

D.P.dL. MINECO: program FIS2015-67367-C2-1-p

E.M.P European Research Council (ERC-StG-MINT 307609)

E.M.P. MINECO: CTQ2014-60541-P

E.G. AMAROUT II fellowship program: grant for transnational mobility (Marie Curie Action, FP7-PEOPLE-2011-COFUND (291803))

P.N, T.K: DFG: GRK 1570 and KO3612/1-1

JML: DFG SFB 689

J.L.L., J.F.R. Marie-Curie-ITN 607904-SPINOGRAPH

**SUPPLEMENTARY INFORMATION**

The supplementary information contains additional details about the synthesis of PbI$_2$ nanosheets, and the optical and TEM characterization of the crystals, additional photoluminescence data and analysis, measurements from the synchrotron, data from additional photodetectors and computational details.

# Supplementary Information: Characterization of highly crystalline lead iodide nanosheets prepared by a room-temperature solution processing

Riccardo Frisenda*[1], Joshua O. Island[2,+], Jose Lado[3], Emerson Giovanelli[1], Patricia Gant[1,4], Philipp Nagler[5], Sebastian Bange[5], John M. Lupton[5], Christian Schüller[5], Aday Molina-Mendoza[6,++], Lucia Aballe[7], Michael Foerster[7], Tobias Korn[5], Miguel Angel Niño[1], David Perez de Lara[1], Emilio M. Pérez[1], Joaquín Fernandéz-Rossier[3] and Andres Castellanos-Gomez*[4]

[1] Instituto Madrileño de Estudios Avanzados en Nanociencia (IMDEA-Nanociencia), Campus de Cantoblanco, E-28049 Madrid, Spain.

[2] Kavli Institute of Nanoscience, Delft University of Technology, Lorentzweg 1, 2628 CJ Delft, The Netherlands.

[3] International Iberian Nanotechnology Laboratory (INL), Av. Mestre Jose Veiga, 4715-330, Braga, Portugal.

[4] Instituto de Ciencia de Materiales de Madrid (ICMM-CSIC), Campus de Cantoblanco, E-28049 Madrid, Spain.

[5] Department of Physics, University of Regensburg, Universitätsstrasse, Regensburg D-93040, Germany.

[6] Departamento de Física de la Materia Condensada, Universidad Autónoma de Madrid, Campus de Cantoblanco, E-28049, Madrid, Spain.

[7] ALBA Synchrotron Light Facility, Carrer de la Llum 2-26, Cerdanyola del Vallés, Barcelona 08290, Spain.

[+] Present address: Department of Physics, University of California, Santa Barbara CA 93106 USA.

[++] Present address: Institute of Photonics, Vienna University of Technology, Gusshausstrasse 27-29, A-1040 Vienna, Austria.

*E-mail: riccardo.frisenda@imdea.org; andres.castellanos@csic.es.

**Section 1 – PbI$_2$ nanosheets production**

Figure S1 shows microscope pictures of four different PDMS substrates where PbI$_2$ crystallites were grown from a hot aqueous solution at different times. As can be seen, the concentration of PbI$_2$ crystals on the surface of the PDMS carrier substrate can vary from sample to sample but overall the crystallization results appear reproducible. Most of the synthesized PbI$_2$ consists of hexagonal crystals with lateral dimensions smaller than 5 µm. One can also find hexagonal platelets with lateral dimensions larger than 20 µm. From the color in the pictures, where the illumination is in transmission mode, one can estimate the thickness of the flake as mentioned in the main text. The pictures of sample 1, 2 and 3 contain a few crystals with a thickness smaller than 25 nm.

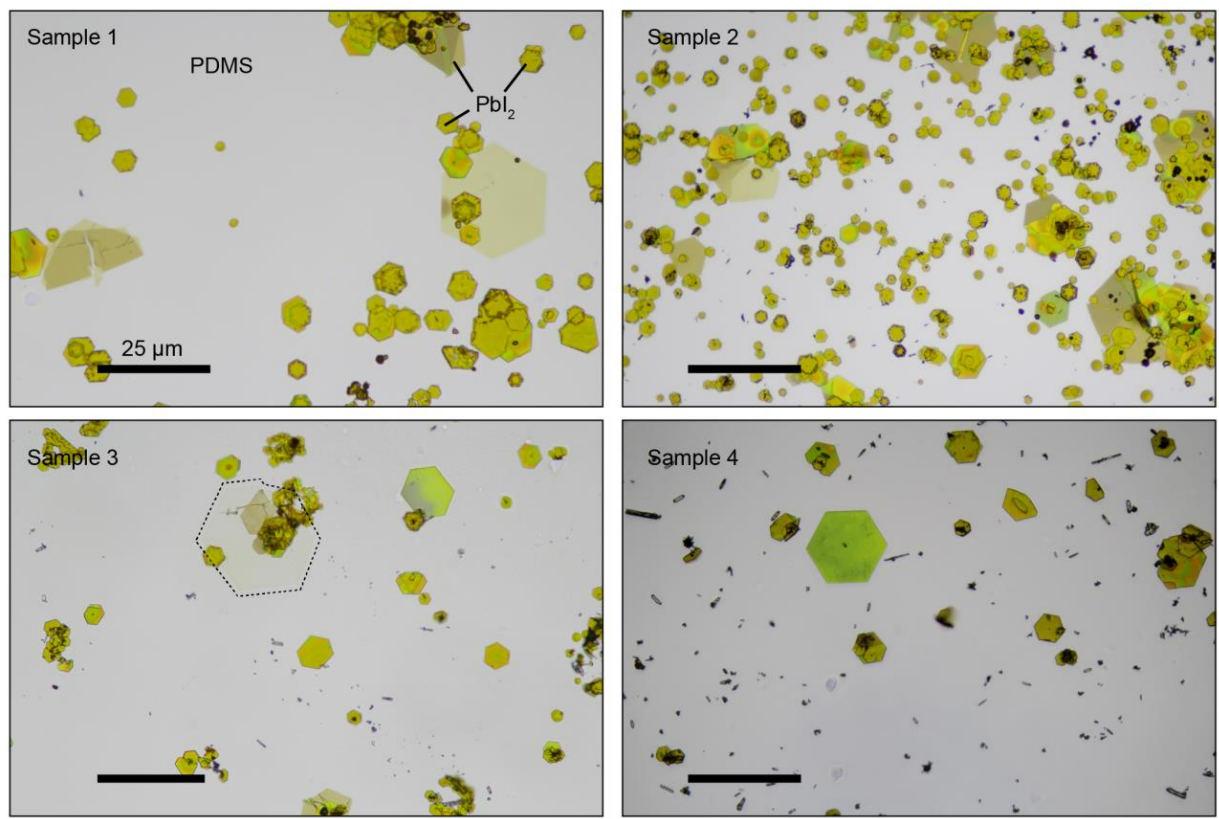

**Figure S1**: Microscope pictures of different samples of PbI$_2$ grown on PDMS substrates. The dashed line in the picture of sample 3 highlights an ultra-thin PbI$_2$ flake.

**Section 2 – PbI$_2$ characterization**

Figure S2a shows microscope pictures of four different PbI$_2$ ultra-thin flakes on PDMS recorded in two different illumination modes. For each flake we recorded the transmittance with a compact CCD spectrometer and Fig. S2b collects the four recorded transmission spectra. The spectra of flakes 2, 3 and 4 each have a clear step in the transmission located at 2.4 eV where the material starts to transmit less light. This energy matches well the value of the direct bandgap found from the different techniques mentioned in the main text. Figure S2c replots the same data as absorbance spectra.

To characterize the crystal structure of the synthesized PbI$_2$ crystals, among the various techniques, we employed TEM measurements that are discussed in Fig. 7 of the main text. To perform a quantitative analysis of the TEM image we extracted the position of each atom and calculated the distance between every pair of atoms, with an automatic algorithm written in Matlab. Figure S3a shows the atoms identified in the bottom part of the image, revealing that the algorithm is able to identify most of the atoms. In Fig. S3b we show a magnification of the previous image where we colored an atom in black. The atoms colored in blue and green are respectively the first- and second-nearest neighbor of the black atom. After extracting the distance between every atomic pair in the image we build a histogram from these values and Figure S3c shows the two lowest-distance peaks of the histograms. These peaks correspond respectively to the distance between first- and second-nearest neighbor atoms and each peak can be well fit to a Gaussian peak. Figure S4 shows TEM images of two additional PbI$_2$ ultrathin flakes.

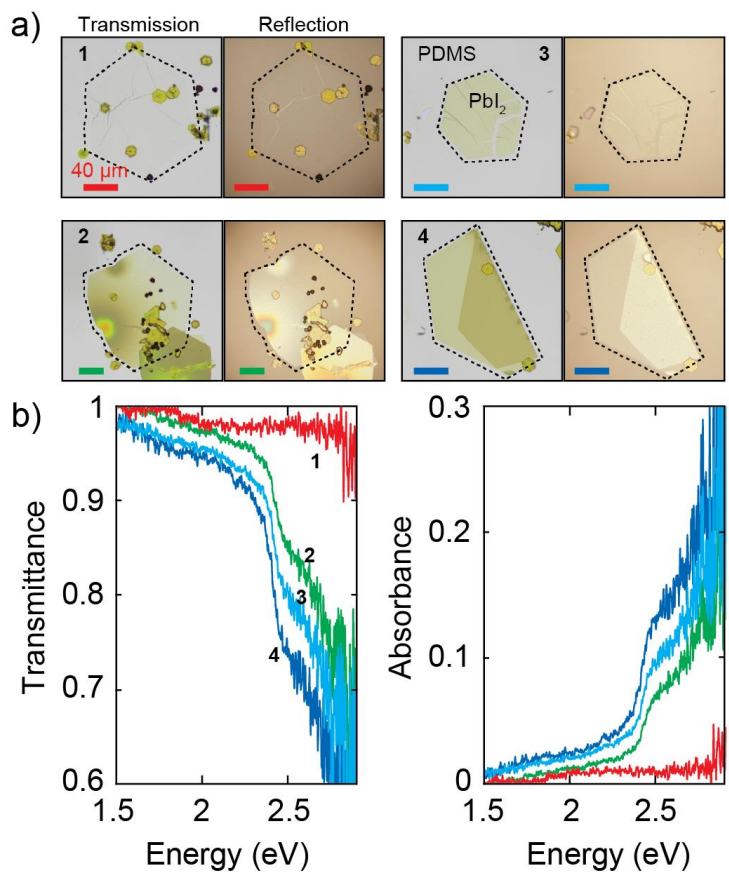

**Figure S2**: a) Microscope optical images of four different PbI$_2$ ultra-thin flakes recorded on PDMS in transmission (left) and reflection (right) illumination mode. The dashed lines are guidelines for the eye. b) Micro-transmittance spectra of the four flakes. c) Absorbance spectra calculated from the transmission spectra in panel (b) by taking the negative base 10 logarithm of the transmission.

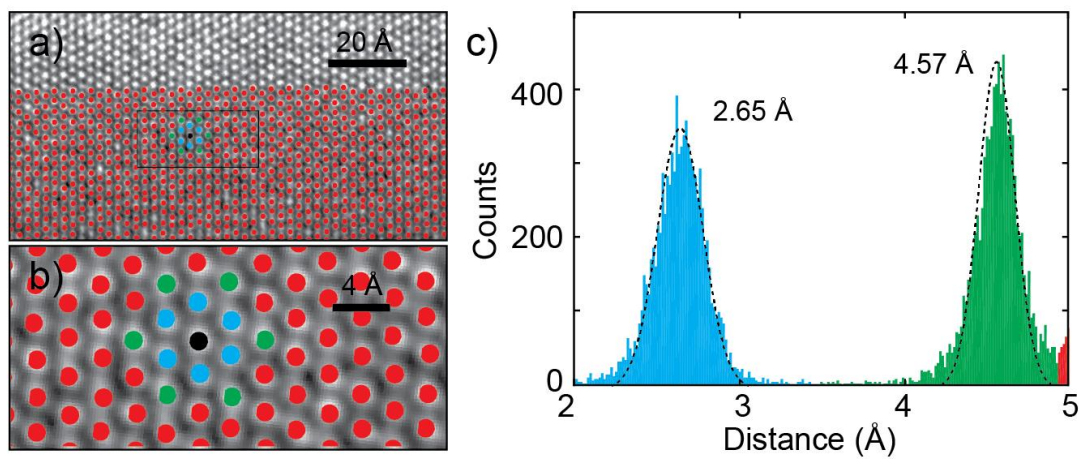

**Figure S3**: a) Zoom of the TEM image of Fig. 7 of the main text. The bottom half of the picture contains colored circles that indicate the atoms recognized by the Matlab algorithm. b) High magnification of the picture in panel (a). Each atom (for example the one indicated by a black circle in the center of the figure) has six nearest neighbor atoms (indicated by the blue circles) and six second-nearest neighbor atoms (green circles). c) Histogram built from all the atomic distances extracted from the TEM image. The two peaks, which can be well fitted by a Gaussian function, are due to the first- and second-nearest neighbor pairs of atoms.

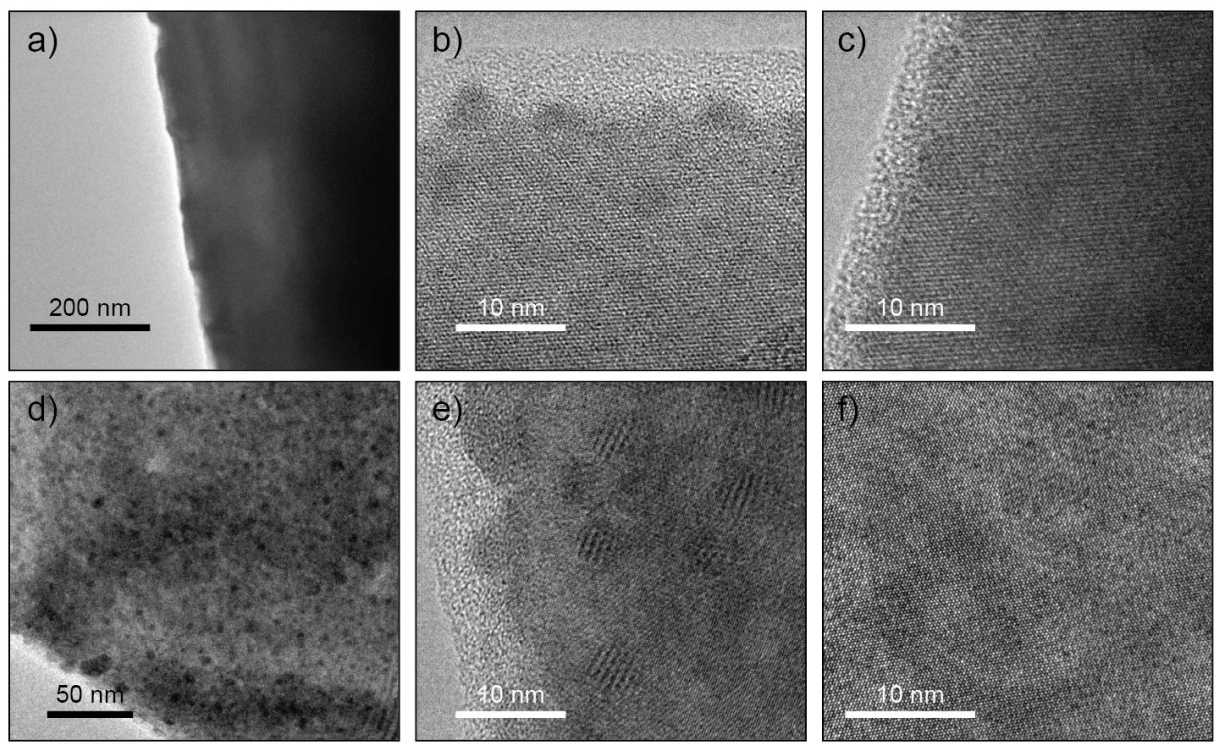

**Figure S4**: a-f) TEM images of two different PbI$_2$ flakes. Panels a and d are low-resolution TEM images while the other panels are high-resolution TEM images where individual Pb atoms are visible.

**Section 3 – Optical contrast of PbI$_2$ on SiO$_2$/Si**

Figure S5a shows a schematic of the optical contrast experiment. By measuring the light reflected from the PbI$_2$ flake and the light reflected from the SiO$_2$/Si substrate one can extract the optical contrast. Figure S4b shows the optical contrast of four PbI$_2$ flakes with different thickness. The experimental spectra can be fit to a multilayer Fresnel equation model which takes into account the reflection and the transmission of light across the different materials (air, PbI$_2$, SiO$_2$ and Si). Each material is described by a complex refraction index ($n_1$, $n_3$ and $n_4$) and the unknowns of the model are the real and the imaginary part of the PbI$_2$ complex refraction index ($n_2 = n - i\kappa$). The thickness of SiO$_2$ is 285 nm while both air and Si are considered semi-infinite.

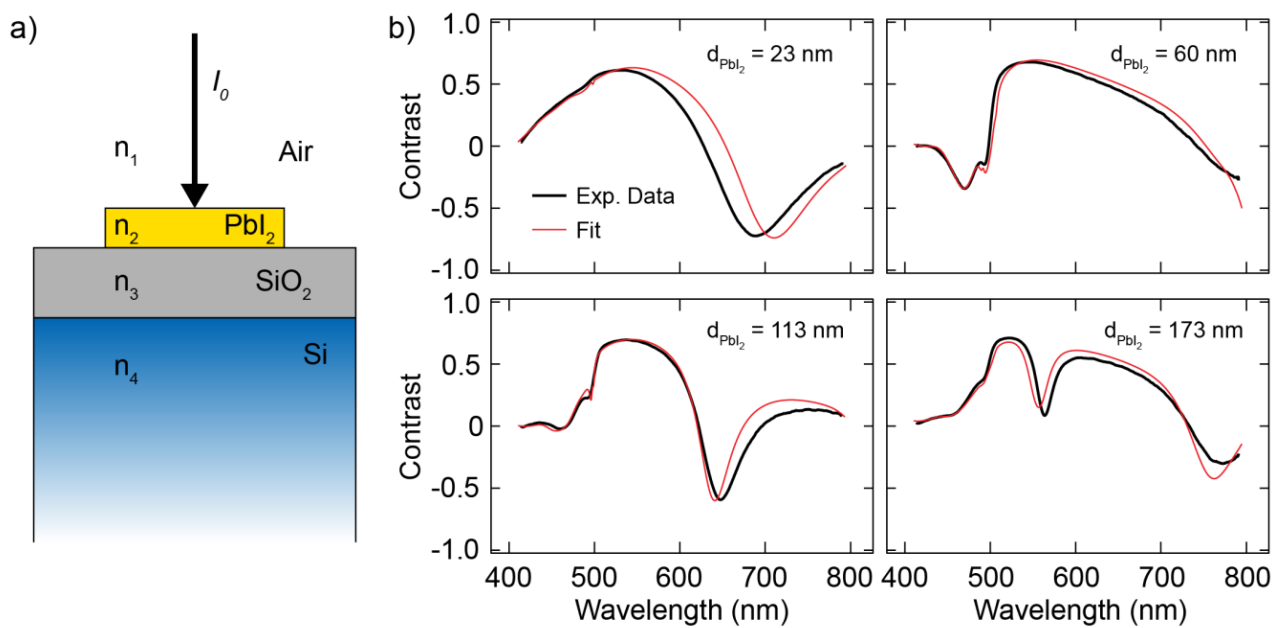

**Figure S5**: Optical contrast of PbI$_2$ flakes with different thickness. The red line is the best fit to the Fresnel multilayer interference model.

**Section 4 – PbI$_2$ photoluminescence**

Figure S6a shows a microscope picture of the PbI$_2$ flake on PDMS used for the photoluminescence experiment discussed in Fig. 8 of the main text. Figure S5b shows the same flake after deterministic transfer onto a SiO$_2$/Si substrate, notice that part of the flake ruptured during the transfer process. AFM measurements were conducted on the sample to extract the thickness of the different regions. Scans of the surface topography were obtained with silicon probes (Nanosensors PPP-NCHR, tip radius <10 nm) using a Park Systems XE-100 in "non-contact" amplitude modulation mode at ~8 nm tip oscillation amplitude (~10 nm free-space oscillation) at ~267 kHz under ambient conditions in a vibration-isolated enclosure. Line scan results are collected in Fig. S6c. We plot in Figure S7a a linear representation of the low temperature PL spectra of the main text. Figure S7b shows the intensity of the PL peaks at 2.49 eV, 2.42 eV and 2.05 eV as a function of the laser incident power plotted in a log-log representation. From the graph it is evident that the intensity of the peaks at 2.49 eV and 2.42 eV follow a power law with a different exponent (>1.3) than the peak at 2.05 eV (whose exponent is 0.8). Upon increasing the temperature the bandgap of PbI$_2$ experiences a red-shift and in Fig. S8 we plot this dependency for the 23 nm thick flake. The solid line in the figure is a fit to the Bose-Einstein model:

$$E_G(T) = E_G(0) - 2a / [\exp(T_0 / T) - 1], \quad (1)$$

where $E_G(T)$ is the bandgap energy at a temperature $T$ and $a$ and $T_0$ are two parameters of the fit which are related respectively to the exciton-phonon interaction strength and to the Debye temperature. From the fit we find that the bandgap at 0 K is 2.50 eV, $a$ = 285 meV and $T_0$ = 606 K.

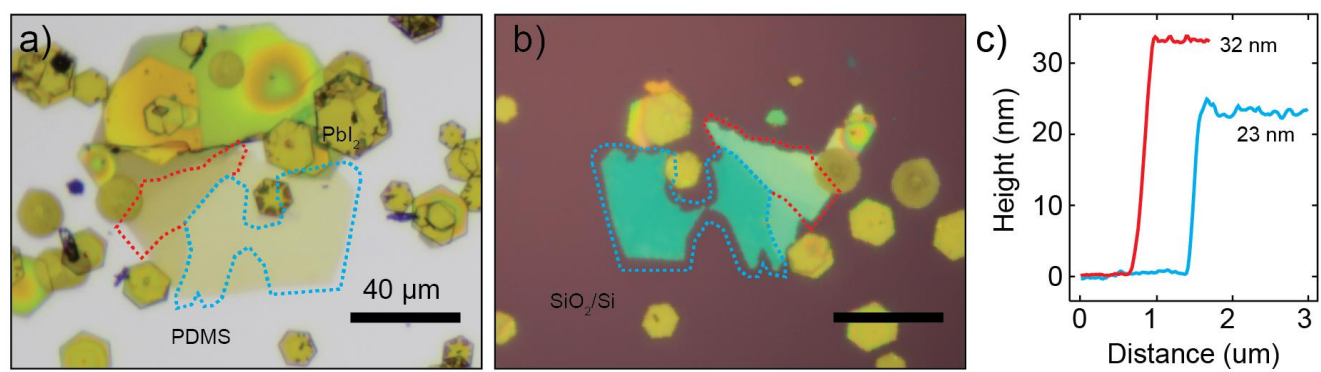

**Figure S6**: a) Optical picture of the flake of PbI$_2$ used for the photoluminescence studies of the main text. b) Optical picture of the same flake shown in (a) after transferring to SiO$_2$/Si substrate. c) AFM profiles of the flake.

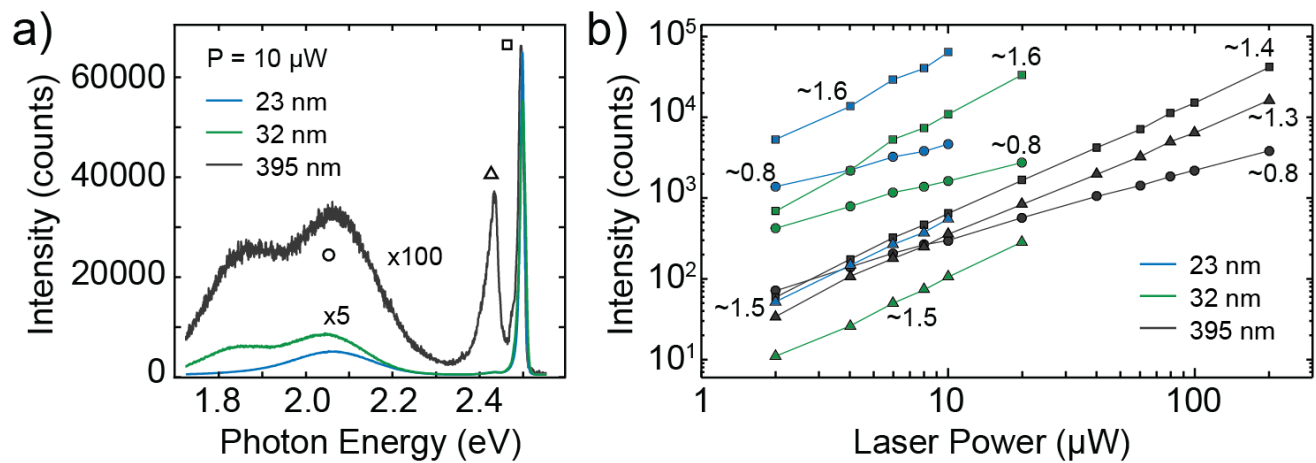

**Figure S7**: a) Photoluminescence spectra for three different PbI$_2$ thicknesses in linear scale. b) Log-log plot of the intensity of the photoluminescence peaks at 2.49 eV (squares), 2.42 eV (triangles) and 2.05 eV (circles) as a function of laser excitation power for the three thicknesses. The numbers indicate the exponent of a power law fit performed on each curve (a value of 1 indicates a linear dependence).

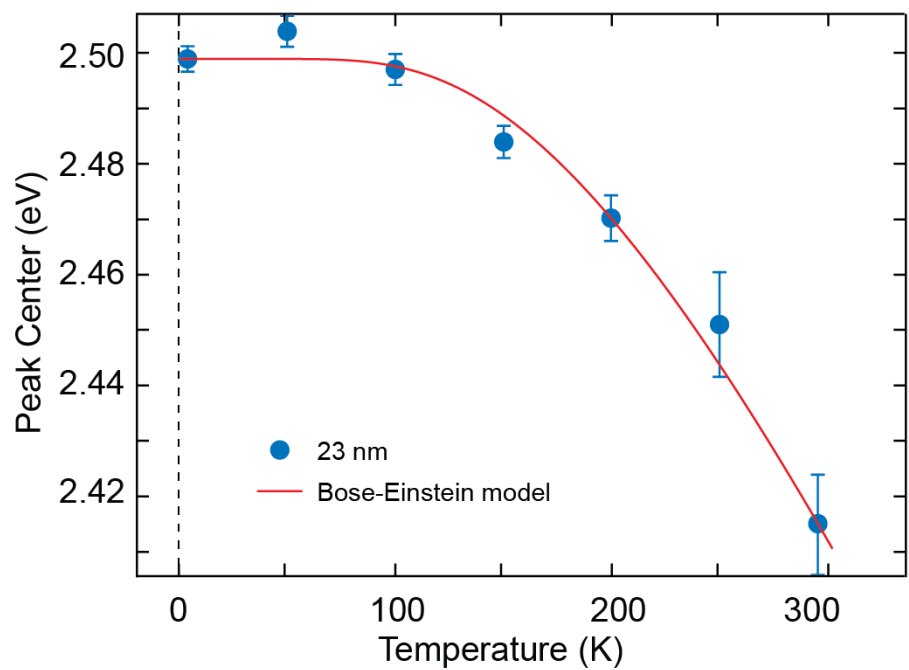

**Figure S8**: a) Energy of the highest-energy photoluminescence peak as a function of temperature. The red line is a fit to the Bose-Einstein model for the temperature dependence of the bandgap.

**Section 5 – Synchrotron measurements**

We performed laterally resolved x-ray photoemission spectroscopy (micro-XPS) at Circe beamline in ALBA Synchrotron Facility [1] on a $PbI_2$ flake transferred to a Pt surface. With LEEM/MEM and micro-LEED we have measured the surface morphology and the structure shown in Fig. S9a-c. We can compare the electron microscopy images showing the morphology of the flakes with the optical images taken at the 2D Materials & Devices Laboratory at IMDEA. The samples are kept 24 hours at room temperature in UHV (at $10^{-8}$ - $10^{-9}$ mbar) in a preparation chamber previous to the experiment, and after that we insert the sample in the measurement chamber ($10^{-10}$ mbar) for the micro-XPS measurements.

We have measured the spin orbit split core levels of iodine I $3d_{3/2}$ - I $3d_{5/2}$ (with photon energy 700 eV) and lead Pb $4f_{5/2}$ - Pb $4f_{7/2}$ (measured with photon energy 230 eV), reported in Fig. S9d. In the region outside the flake we do not detect I nor Pb signal as expected; inside the flake the I 3d doublet is separated by 11.5 eV and the Pb 4f doublet is separated by 4.9 eV in good accordance both with theory and experiments. The I $3d_{5/2}$ core level has a binding energy of 619 eV similar to the values reported in literature [2-4]. The Pb 4f shows a two-component state, having the Pb$4f_{7/2}$ energies at 138.9 eV, a value consistent with the literature for the Pb(II) state, and a second component at 136.85 eV corresponding to a neutral metallic state.

Additionally, we have observed desorption of iodine from the $PbI_2$ with time when the sample is placed in vacuum and with x-ray irradiation, as well as with annealing at a temperature of 150 °C. This desorption is responsible for a stoichiometry of $PbI_{0.93}$ due to an excess of metallic lead, and it appears that there is a dependence of the I/Pb ratio on the thickness of the flake, with a decreasing amount of iodine relative to Pb for the thinner regions, indicating that in the thinner part of the flake, I is desorbing faster than in the thicker part. See Fig. S10a for a comparison of the core levels Pb 4f and I 3d measured before and after annealing at 150 °C. The lead core levels present a shift to lower binding energy, typical of metallic Pb, while the iodine core levels disappear completely. This conversion of $PbI_2$ to Pb is also apparent from the optical pictures in Figs. 10b-c. The dramatic change in color and contrast in the optical images indicates a large modification of the material optical properties.

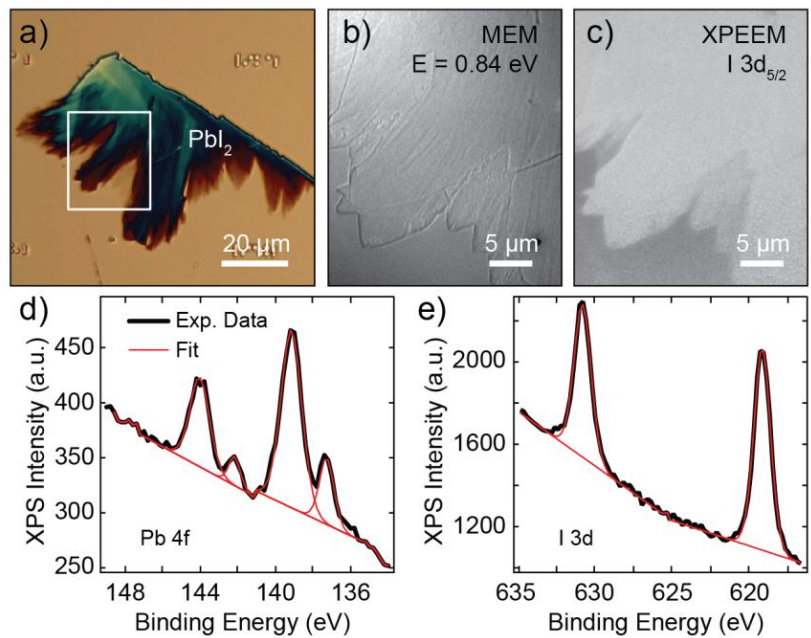

**Figure S9**: a) Optical image of a $PbI_2$ flake transferred on a platinum substrate for synchrotron studies. b-c) Mirror electron microscopy image and X-ray Photoemission image of the region highlighted by the white rectangle in panel (a). d-e) Core levels of Pb $4f_{5/2}$ - Pb $4f_{7/2}$ and I $3d_{3/2}$ - I $3d_{5/2}$.

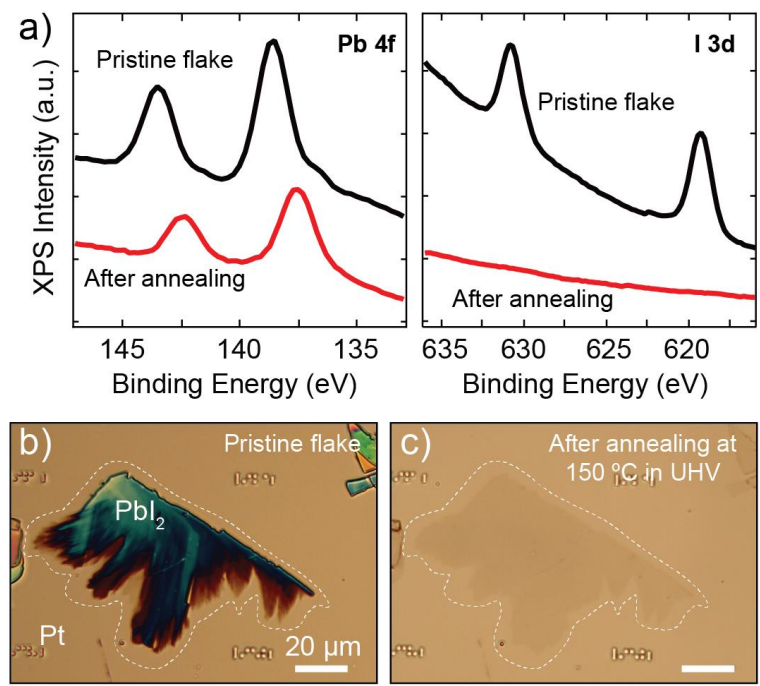

**Figure S10**: a) XPS core levels of Pb 4f and I 3d measured on a pristine PbI$_2$ flake (black line) and after annealing at 150 °C in ultra-high vacuum (red line). b) Optical image of a PbI$_2$ flake transferred on a platinum substrate. c) Same flake after annealing at 150 °C in ultra-high vacuum. The change in color reflects the change in the chemical composition after desorption of iodine atoms.

### Section 6 – Additional photodetectors

The thickness of the PbI$_2$ crystal used to fabricate the photodetector discussed in the main text, device A depicted in Figure S11a, was estimated with AFM measurements. Figure S11b shows the topography of the device which displays a good spatial homogeneity. From the line profile in figure S10c we extract 15 nm as the thickness of the flake. Figure S12 shows the photocurrent generated in device A at λ = 405 nm as a function of the optical power. The black line is a power law fit to the data from which we extract 0.7 as the exponent, indicating a sub-linear dependence of the photocurrent on the power.

Figure S13 shows an additional PbI$_2$ photodetector (device B) fabricated on gold pre-patterned electrodes on 285 nm SiO$_2$/Si substrate. Figure S13a shows the flake on PDMS prior to transfer, while Figure S13b shows the fabricated device. Figures S13c and S13d show the electrical characterization of the device in dark and under illumination. The responsivity of this device is comparable to the one of device A. We fabri-cated in total four different photodetectors and we found similar performances in all of them as shown in the comparison plot of Figure S14. The values of the responsivity and the energy dependence are similar in all four cases. Figure S16 shows the aging of another photodetector, after 15 days of fabrication the responsivity decreased of 90% in respect to the as-fabricated value.

We also fabricated a photodetector on a flexible and transparent polycarbonate substrate. Figure S15a shows an optical picture of the device recorded in transmission illumination mode. Figure S15b shows the current-voltage characteristics of the device in dark and under illumination. The extracted responsivity R = 0.35 mA/W is comparable to the one found in the devices built on $SiO_2$/Si.

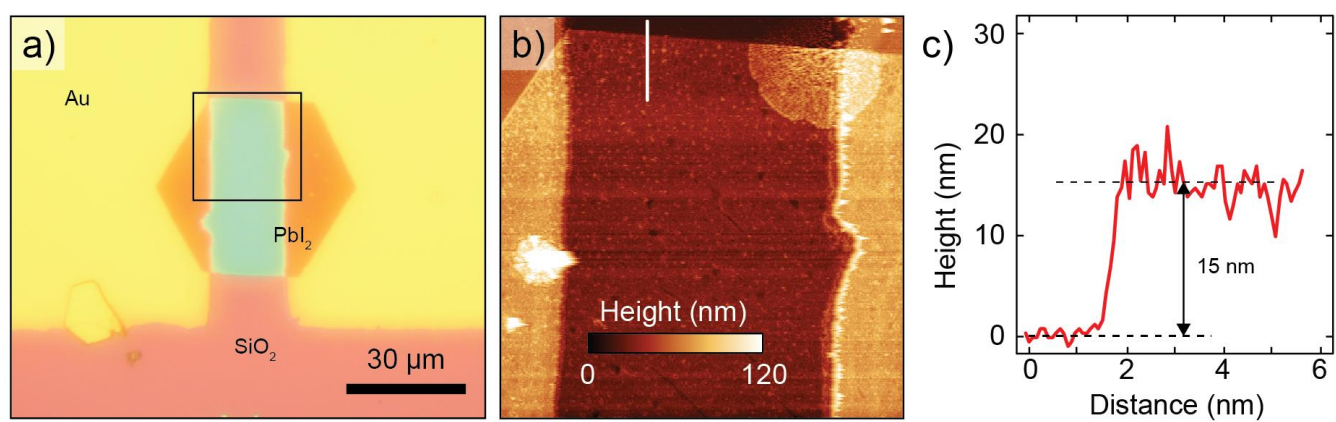

**Figure S11**: a) Optical picture of device A, which is the photodetector discussed in the main text. b) AFM topographic image of the region indicated with a black square in panel (a). c) Height profile extracted at the position indicated by the white line in panel (b).

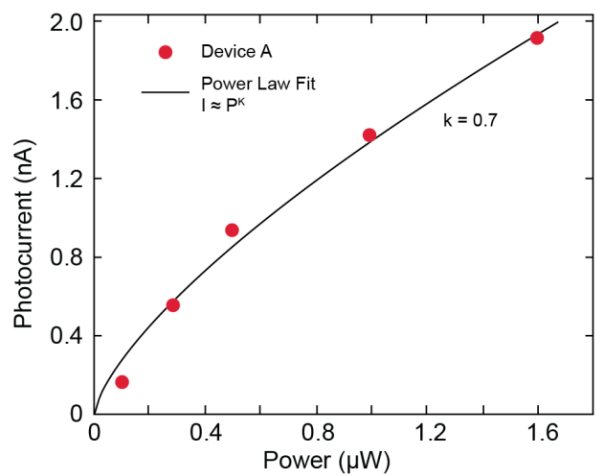

**Figure S12**: Photocurrent as a function of the optical power for device A at λ = 405 nm and bias voltage 10 V. The line is a power law fit with exponent 0.7.

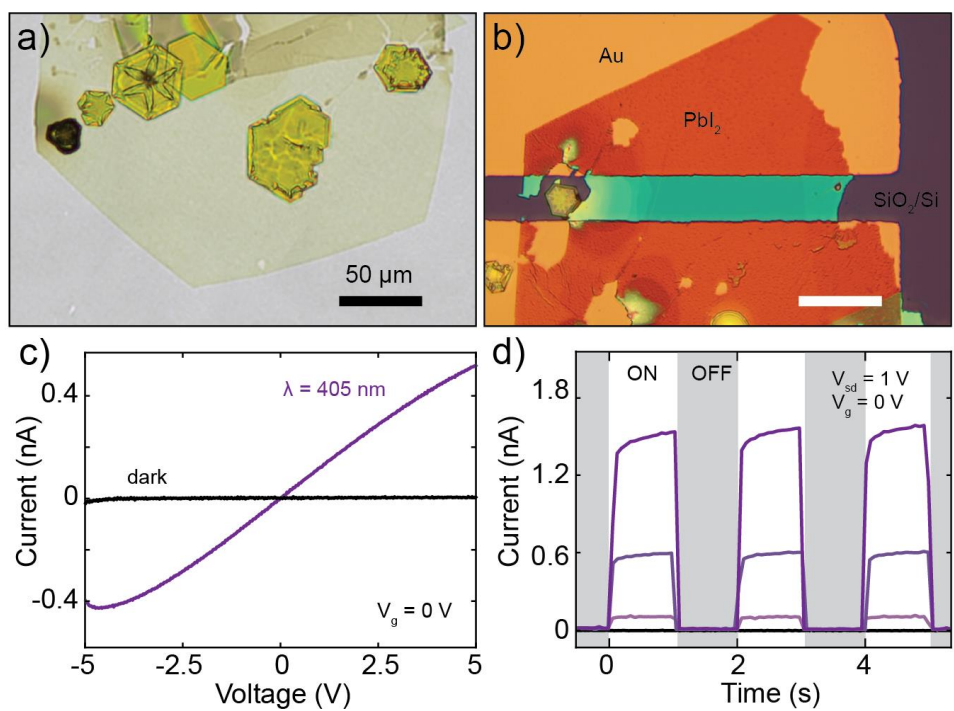

**Figure S13**: a-b) Optical pictures of a PbI$_2$ flake on PDMS and on SiO$_2$/Si after deterministic transfer. This photodetector is sample B. c) Current-voltage characteristics of the device in dark and under global illumination at 405 nm.

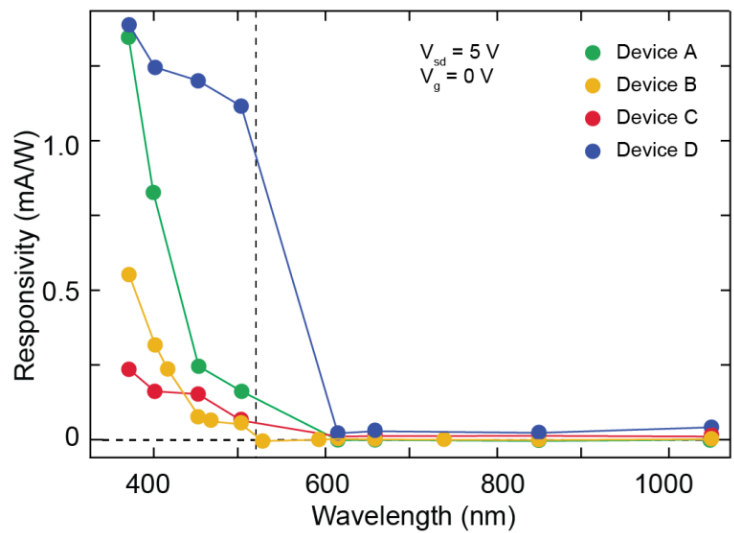

**Figure S14**: Wavelength-resolved responsivity of four different PbI$_2$ devices recorded at 5 V.

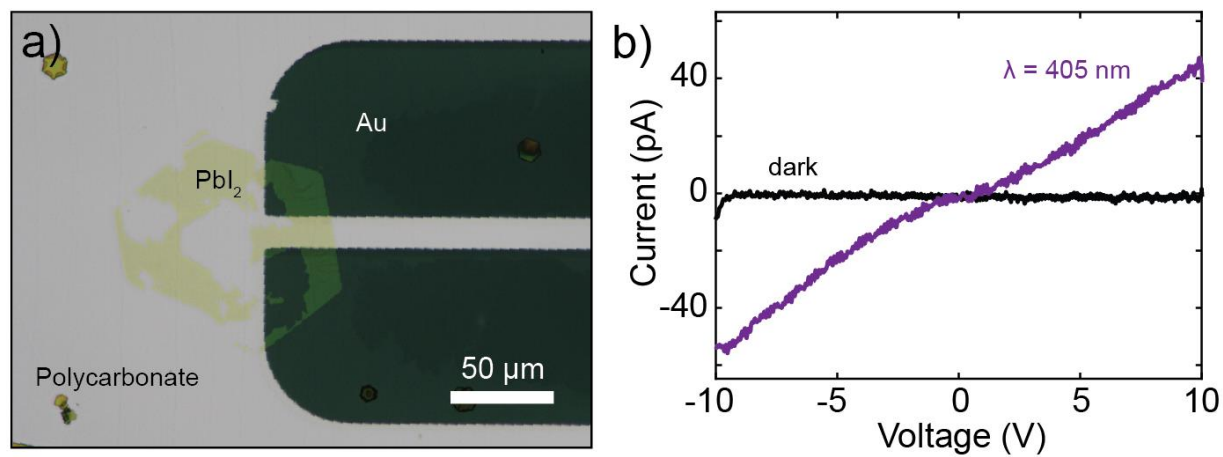

**Figure S15**: a) Optical picture (recorded in transmission illumination mode) of a flexible photodetector composed by a thin crystal of PbI$_2$ transferred between pre-patterned gold electrodes on polycarbonate. b) Current-voltage characteristics of the device in dark and under global illumination at 405 nm. The calculated responsivity is 0.35 mA/W at a voltage of 5 V, which is a value comparable to the devices fabricated onto 285 nm SiO$_2$/Si substrate.

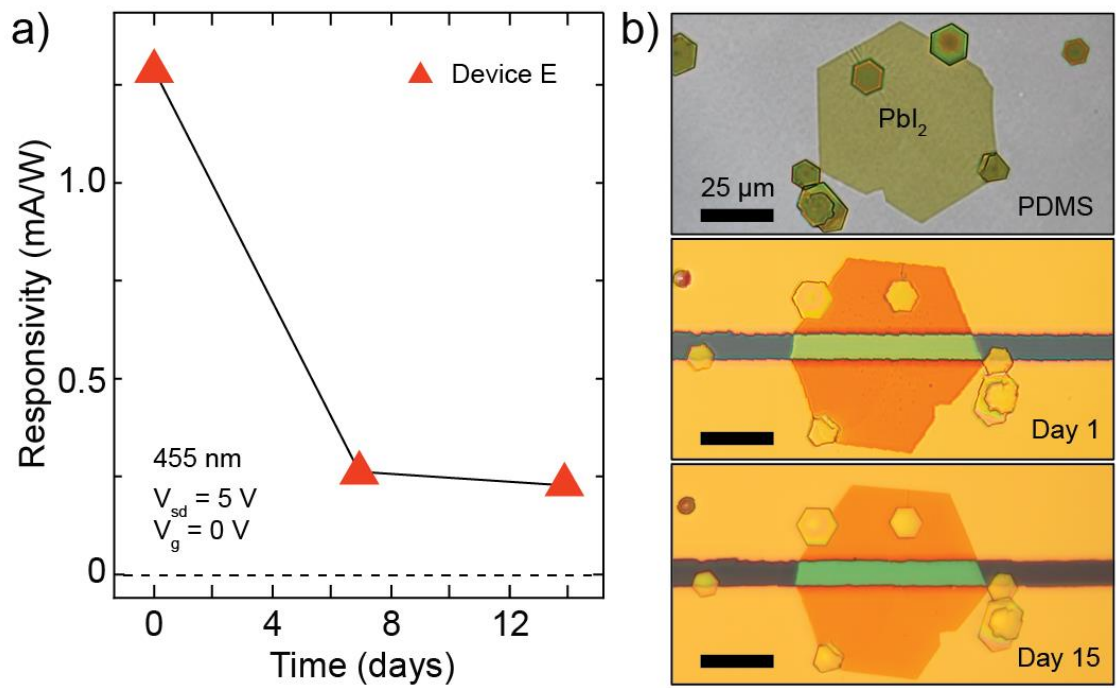

**Figure S16**: a) Responsivity of device E recorded at 5 V with excitation wavelength of 455 nm recorded for the pristine device and after 7 and 14 days. b) Optical picture of the PbI$_2$ flake isolated on PDMS used to fabricate the photodetector (top) and of the fabricated photodetector after 1 day and after 15 days (bottom).

### Section 7 – Computational methods

Density functional theory calculations were performed using the computational packages Quantum Espresso [5] (QE) and Elk [6]. Geometry optimizations were performed with Quantum Espresso, using PAW pseudopotentials [7] PBE exchange correlation functional [8] and Van der Waals Grimme correction [9] for the bulk compound. With the relaxed structures, all-electron LAPW calculations with Elk were performed with spin-orbit coupling and the Tran Blaha modified Becke Johnson potential [10] (TBmBJ) in order to obtain a reliable band gap for bulk. The TBmBJ constant was calculated selfconsistently for the bulk structure, whereas for the monolayer was fixed to the bulk value in order to avoid spurious effects coming from the vacuum spacing of the unit cell in the monolayer geometry.

In order to obtain as function of the PbI$_2$ thickness, we first obtained a reliable tight binding model for the bulk structure by means of Wannierization, using Wannier90 [11], of the QE band structure, calculated with spin-orbit coupling, PBE exchange correlation functional and norm conserving pseudopotentials. The PBE conduction band eigenvalues were shifted before the Wannierization to reproduce the TBmBJ bulk band gap. With the previous Wannier Hamiltonian for bulk $PbI_2$, we build layers with different thickness and we calculate their direct band gap in the tight binding model. It is worth to remark that the previous top-down Wannier-based procedure predicts a band-gap for the monolayer of 2.86 eV, whereas the fully *ab initio* band gap for monolayer is 2.89 eV, yielding almost an identical result. Figure S17 shows the band structure of bulk $PbI_2$ calculated from *ab initio* with and without contributions from the spin-orbit coupling.

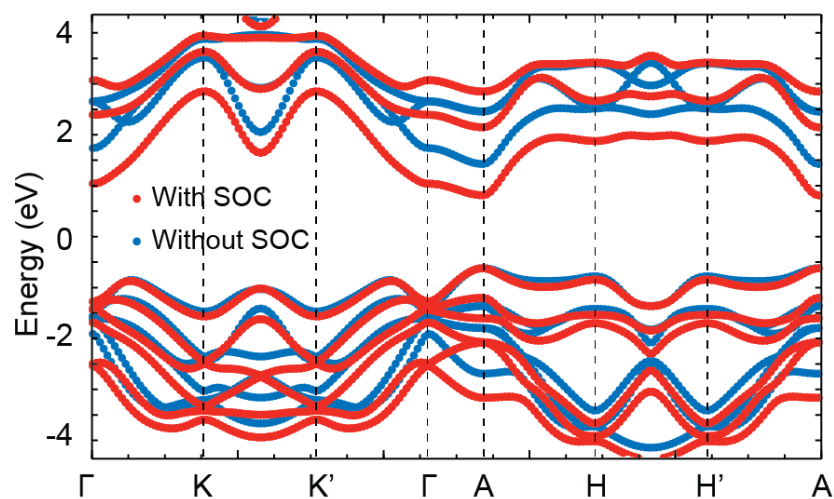

**Figure S17**: Band structure of bulk $PbI_2$ with (red) and without (blue) spin-orbit coupling.

## Section 8 – References